\documentclass[pre,preprint,floatfix,amsfonts,showpacs]{revtex4}
\usepackage{graphicx,graphics,color,epsfig}
\usepackage{bm}
\usepackage{amsmath}
\usepackage{amssymb}
\begin{document}

\title{Quasispecies Theory for Evolution of Modularity}
\author{Jeong-Man Park, Liang Ren Niestemski, and Michael W. Deem}
\affiliation{Department of Physics \& Astronomy,
Rice University, Houston, TX 77005--1892, USA\\
Department of Physical and Biological Science,
Western New England University
Springfield MA, 01119\\
Department of Physics, The Catholic University of Korea, Bucheon
420-743, Korea
}

\begin{abstract}
Biological systems are modular, and this modularity evolves over time and
in different environments.  A number of observations have been made of
increased modularity in biological systems under increased environmental
pressure.  We here develop a quasispecies theory for the dynamics of modularity 
in populations of these systems.  
We show how the steady-state fitness in a randomly changing
environment can be computed.
We derive a fluctuation dissipation relation
for the rate of change of modularity and use it to 
derive a relationship between rate of environmental changes
and rate of growth of modularity.
We also find a principle of least action for the
evolved modularity at steady state.
Finally, we compare our predictions to simulations of protein
evolution and find them to be consistent.
\end{abstract}

\pacs{87.10.-e, 87.15.A-, 87.23.Kg}

\maketitle

\section{Introduction}

  Biological systems have long been recognized to be modular.
  In 1942 Waddington presented his now classic description of a canalized 
landscape for development, in which minor perturbations do
not disrupt the function of developmental modules
\cite{Waddington}.  In 1961 H.\ A.\ Simon described how biological systems are more efficiently evolved
and are more stable if they are modular \cite{Simon}.  A seminal paper by Hartwell \emph{et al.}\ firmly established the
concept of modularity in cell biology \cite{Hartwell1999}.  Systems biology has since provided a wealth of examples
of modular cellular circuits, including metabolic circuits \cite{Ravasz2002,Callahan2009} and modules
on different scales, \emph{i.e.}\ modules of modules \cite{daSilva2008}.  Protein-Protein interaction networks have been observed to be modular \cite{Spirin2003,Gavin2006,vonMering2003}.  Ecological food webs have been found to be
modular  \cite{Krause2003}.
The gene regulatory network of the developmental pathway exhibits modules \cite{Raff2000a,Wagner1996}, and the
developmental pathway is modular  \cite{Klingenberg2008}.  Modules have even been found in physiology, specifically
in spatial correlations of brain activity 
\cite{Meunier2009,Chavez2010}.

The modularity of a biological system can change over time.  There are
a number of demonstrations of the evolution of 
modularity in biological systems.  For example,
the modularity of
the protein-protein interaction network significantly increases when 
yeast is exposed to heat shock
\cite{Csermely}, and the modularity of the protein-protein networks in both yeast and E.\ coli
appears to have increased over evolutionary time \cite{jun2}.
Additionally,
food webs in low-energy, stressful
environments are more modular than those in plentiful environments \cite{Dirk},
arid ecologies are more modular during droughts \cite{Rietkerk}, and
foraging of sea otters is more modular when food is limiting \cite{Tinker}.
Other complex dynamical systems exhibit time-dependent modularity as well.
The modularity of social networks changes over time: 
stock brokers instant messaging networks are more modular under stressful market conditions \cite{Uzzi}, and
socio-economic community overlap decreases with increasing stress \cite{Estrada}.
Modularity of financial networks changes over time: the modularity
of the world trade network has decreased over the last 40 years, leading to increased susceptibility
to recessionary shocks \cite{He2010a}, and
increased modularity has been suggested as a way to increase the robustness and adaptability
of the banking system \cite{May}.
Much of the research on
modularity has suggested that gene duplication, horizontal gene
transfer, and changes in the total number of connections may all play
a role in the evolution of modularity \cite{Hallinan2004a,Rainey2004,jun}.
 
In an effort to proceed further with these observations, we here 
present a quasispecies theory for the evolutionary dynamics of modularity.
This analytical theory complements numerical models that have
investigated the dynamics of modularity \cite{Lipson2002,Alon2005,Alon2007,jun}.
We assume that modularity can be quantified in the system under study.  
We further assume
that modularity is a good order parameter to describe the state of the system.
That is, we project the dynamics onto the slow
mode of modularity, $M$.
In section II we introduce the quasispecies description
for the dynamics of modularity.  
The details of the sequence level evolutionary dynamics
are what, when projected out, define the fitness function $f(m)$ introduced
in this section.
In section III we show how the steady-state fitness
in a randomly changing environment can be computed
from the time-dependent average fitness starting from
random initial conditions.
In section IV we derive a fluctuation dissipation
theory for the dynamics of modularity.
In section V we derive a relationship between rate
of environmental change and rate of growth of modularity.
In section VI we find the evolved, steady-state
value of modularity by a principle of least action.
In section VII we compare some of the predictions
to simulations of protein evolution.
We conclude in section VIII.

\section{The Quasispecies Theory for Dynamics of Modularity}

Quasispecies theory captures the basic aspects of mutation and evolutionary
selection in large, evolving populations \cite{ei71,ck70}.  These models have
been widely used in the physics literature to describe evolutionary
biology \cite{Krug}.  A series of papers showed how these models could be
solved in the steady-state limit, first by a mapping to an inhomogeneous
Ising model \cite{Leuthausser,Tarazona,bb97,bb98,sh04a} and later by solution
with functional integral techniques \cite{Peliti2002,Deem2006,Park06}.
A Hamilton-Jacobi approach has been used to derive dynamical predictions
in these models \cite{de08}.  Quasispecies theory has been extended
to larger alphabets \cite{Munoz3} and to
describe the effects of horizontal gene transfer
\cite{Cohen05,Park,Munoz2}
and finite populations \cite{Park3,Park4}.

We here develop quasispecies theory for the dynamics of modularity.
We consider a population of systems,
where each system is characterized by a specific connection matrix, from
which the modularity can be calculated.  Evolution occurs
within each system by mechanisms such as point mutation or
horizontal gene transfer. Horizontal gene transfer is not
allowed between systems, because such events would violate
the assumption that the fitness of each system depends only
on the modularity of that system.
Competition occurs both within and between systems.  The 
evolutionary dynamics of this population of 
systems is fully specified by the rate at which each system
reproduces, $f$, termed ``fitness,'' and the rate at which changes of
modularity arise, $\mu$.  Since the state of each system is specified
by the slow modularity variable, $M$, the fitness is a function of the modularity, $f=f(M)$.
The $f(M)$
function is from a detailed calculation, numerical simulation, or 
experimental observation of the competitive evolutionary dynamics
within each system with a given value of modularity.
Thus, the rate at which a system with modularity $M$ replicates, $f(M)$,
is an input to the theory to be derived here.
The present theory predicts how modularity
in the  population of systems will evolve, given the replication rates and
mutation rates.

The fitness function $f(M)$ fundamentally 
characterizes an evolving network.   
With this $f(M)$, the dynamics of modularity can be calculated.
For example, the $f(M)$ could be deduced for the
evolution of the protein-protein interaction network
in \emph{E.\ coli}, showing the evolutionary advantage of
modularity for this system  \cite{jun2}.
The $f(M)$ is the driving force for
spontaneous emergence of modularity in a protein network \cite{jun}.
The $f(M)$ quantifies the benefit of modularity to
a system, and we will show that modularity evolves
to a finite modularity at steady state in a population of systems.

Modularity is defined on a network of nodes and edges.  Thus, the
fundamental object describing each system is the connection
 matrix, with the $ij$ element of the connection
matrix representing the value of edge $ij$.
 The connection matrix gives the links
between the nodes of the network.  
For example, in the protein-protein interaction network, the nodes are the proteins
and the links tell one whether protein $i$ interacts with protein $j$.
Modularity of each system is calculated directly from the connection
matrix of that system, and rearrangement of the connections within 
this matrix changes the modularity of a given system.

  The connection matrix, $\Delta_{ij}$, is
a binary matrix that
denotes whether nodes $i$ and $j$ 
interact ($\Delta_{ij}=1$) or not ($\Delta_{ij}=0$).
The detailed dynamics of the system may well have
non-trivial couplings between nodes \cite{jun}, and the
connection matrix is the projection of the non-zero couplings.
We allow each node to be connected to $C$ other nodes on average.
The number of nodes is denoted by $L$.  
Rearrangement of the entries within this matrix changes the modularity of the
matrix.
For simplicity, we assume that the modules which form are of size $l$.
There are two ways to view the fixed partitioning that we consider.
First, this partitioning results from modularity that is induced
by horizontal gene transfer of segments with fixed length $l$, as was
previously shown \cite{jun,jun2}.
Second, biological modules are often of roughly fixed size, so it is not
too much of a simplification to say the module size is constant
for all modules.  A fixed partitioning is a subset of all possibilities;
in this work, we consider only this fixed partitioning.
Thus a modular system 
will have an
 excess of connections along the $l \times l$ block diagonals of the 
connection matrix.  In other words, the probability of a connection is
$C_0/L$ outside the block diagonals
 when $ \lfloor i/l \rfloor \ne \lfloor j/l \rfloor$
and $C_1/L$ inside the block diagonals
when $ \lfloor i/l \rfloor = \lfloor  j/l \rfloor$, 
with  $C = C_0 + (C_1 - C_0)l/L$.
Modularity is defined by the excess of connections in the
block diagonals, over that observed outside the
block diagonals: $M = (C_1 - C_0) l / (L C)$.

Modularity changes because the entries in the connection matrix change.
There are several possible models for how the connection matrix
may reorganize.  We here consider the model in which connections
may independently reorganize.  This model is biologically appropriate when
connections between nodes are governed by independent pieces of
structure in each node.  We are not specifically considering ``hub'' nodes that connect
to a very large number of other nodes.  A model of this effect would be
hierarchical.  We are here considering one level of this hierarchy in the
present model.
Thus, we here consider a simple model in which each of these connections
has a rate $\mu$ to rewire.
That is, we define $\mu$ to be 
the rate at which any given
$1$ in the $\Delta$ matrix hops to another random location.
In a typical biological system
there are a finite number of connections per site, even for a large
matrix, and so we consider the limit of $C$ finite and $L$ large, i.e.\ a dilute
matrix of connections.
Thus, the entries in the connection matrix each have rate $\mu $
to independently move to a new position in the connection matrix, and
collisions between connections do not significantly affect the
dynamics in the dilute limit.

When the population of systems is large, the
probability distribution to have a connection
matrix with modularity $m$ obeys (see Appendix A)
\begin{eqnarray}
\frac{d P_m(t')}{d t'} &=& 
L [f (m)- 
\langle f \rangle] P_m(t') + 
\mu C l
 \left[ (1-m) 
\left(1 - \frac{l}{L}
\right)
 +\frac{1}{L C}
\right]
P_{m - 1/ [(L-l)C]}(t')
\nonumber \\ && +
\mu C
\left(L - l \right)
\left[
m + (1-m) \frac{l}{L}
+ \frac{1}{L C}
\right]
 P_{m + 1/[(L-l) C]}(t')
\nonumber \\ && 
- \mu  C
\left(L -l  \right)
\left(
m  + 2 (1-m)  \frac{l}{L}
\right)
 P_{m}(t')
\label{10}
\end{eqnarray}
where $m$ takes values $-l / (L-l), (-l+1/C)/(L-l), (-l+2/C)/(L-l), \ldots, 1$.
The average fitness is given by
\begin{equation}
\langle f(t) \rangle = \sum_m  
f (m)
P_m(t) 
\label{3b}
\end{equation}
The average modularity as a function of time is given by
$M(t) = \sum_m m P_m(t)$.

\section{The Steady-State Fitness in a Randomly Fluctuating Environment}

We here consider how to describe the effect of environmental change
on the evolution of modularity.  We characterize the environmental
changes by their magnitude and frequency.  We denote the
magnitude of environmental change by $p$.  If $p=0$, the environment
does not change at all, and if $p=1$, the environment is
completely different before and after the change.
Although the environmental change is random, on
average a fraction $p$ of the environment's effect on
the fitness of the system is modified by the change.
This model is used to describe evolution of influenza viruses,
where $p$ is defined as above \cite{Deem,Sun}.  In application
to data on influenza vaccines, $p$ 
is termed $p_{\rm epitope}$ and serves as an
accurate order parameter to characterize how effective a vaccine
against one strain will be in protecting against another strain
that is distance $p_{\rm epitope}$ away \cite{Enrique,Gupta2006,Zhou2006}.
Here we consider these environmental changes to occur with a frequency,
which we denote by $1/T$. In particular, we consider that the environmental changes occur
every $T$ timesteps.  This characterization of environmental change
by magnitude and frequency, $p$ and $1/T$, has been
used extensively in the past \cite{Earl,jun,jun2,Dirk,He2010a}.

\begin{figure}[t!]
\begin{center}
\includegraphics[width=2in,clip=]{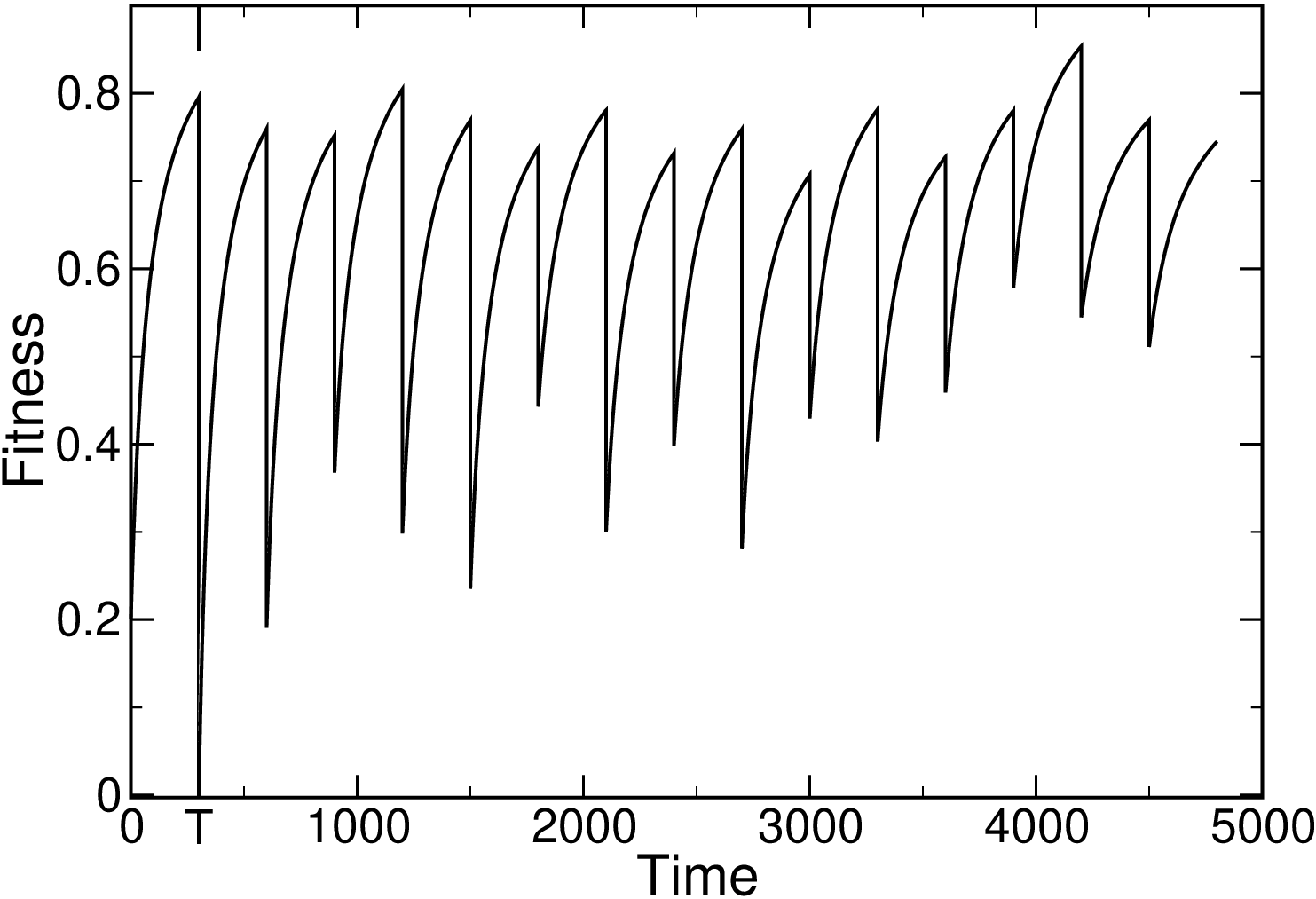}
\includegraphics[width=2in,clip=]{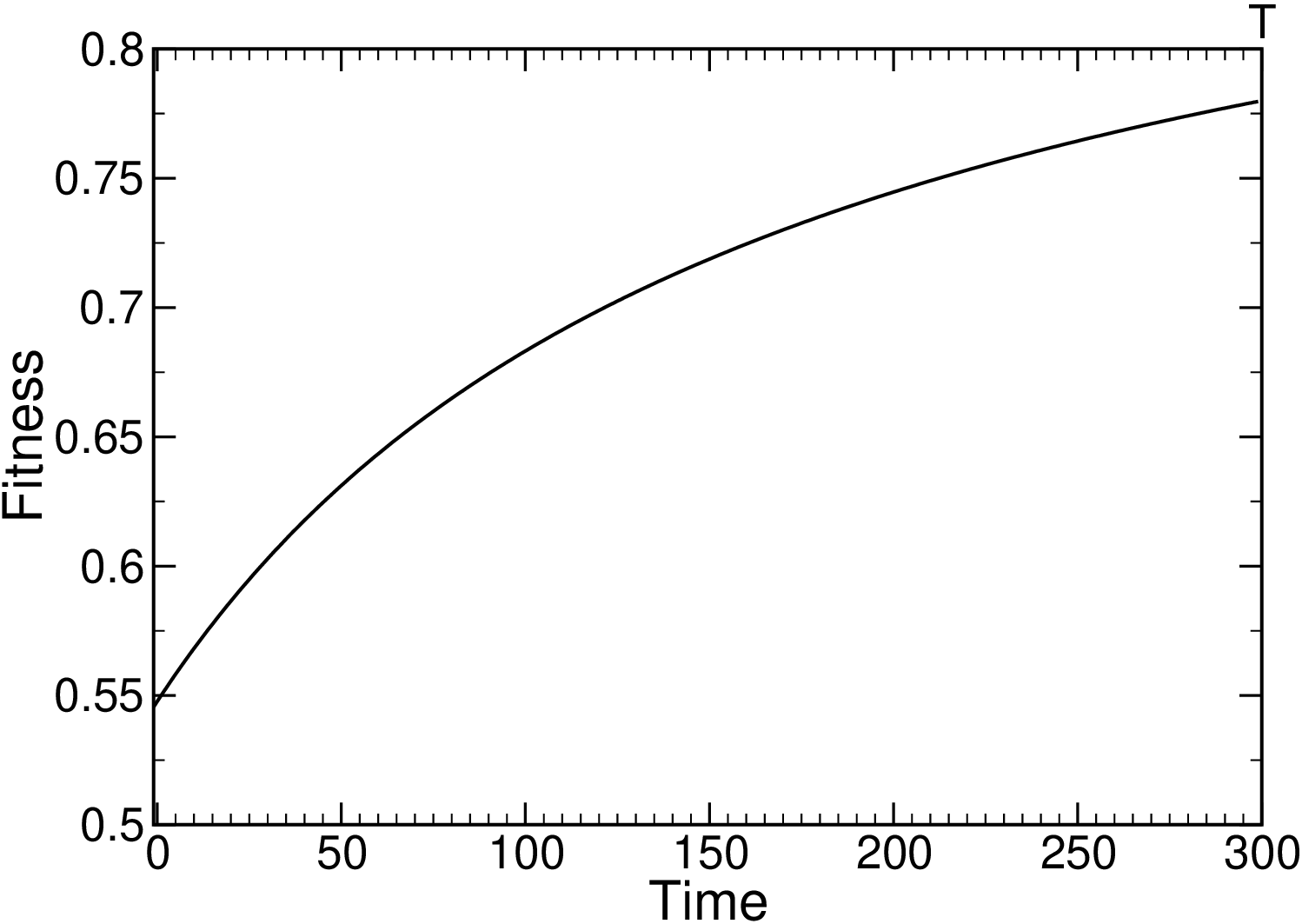}
\includegraphics[width=2in,clip=]{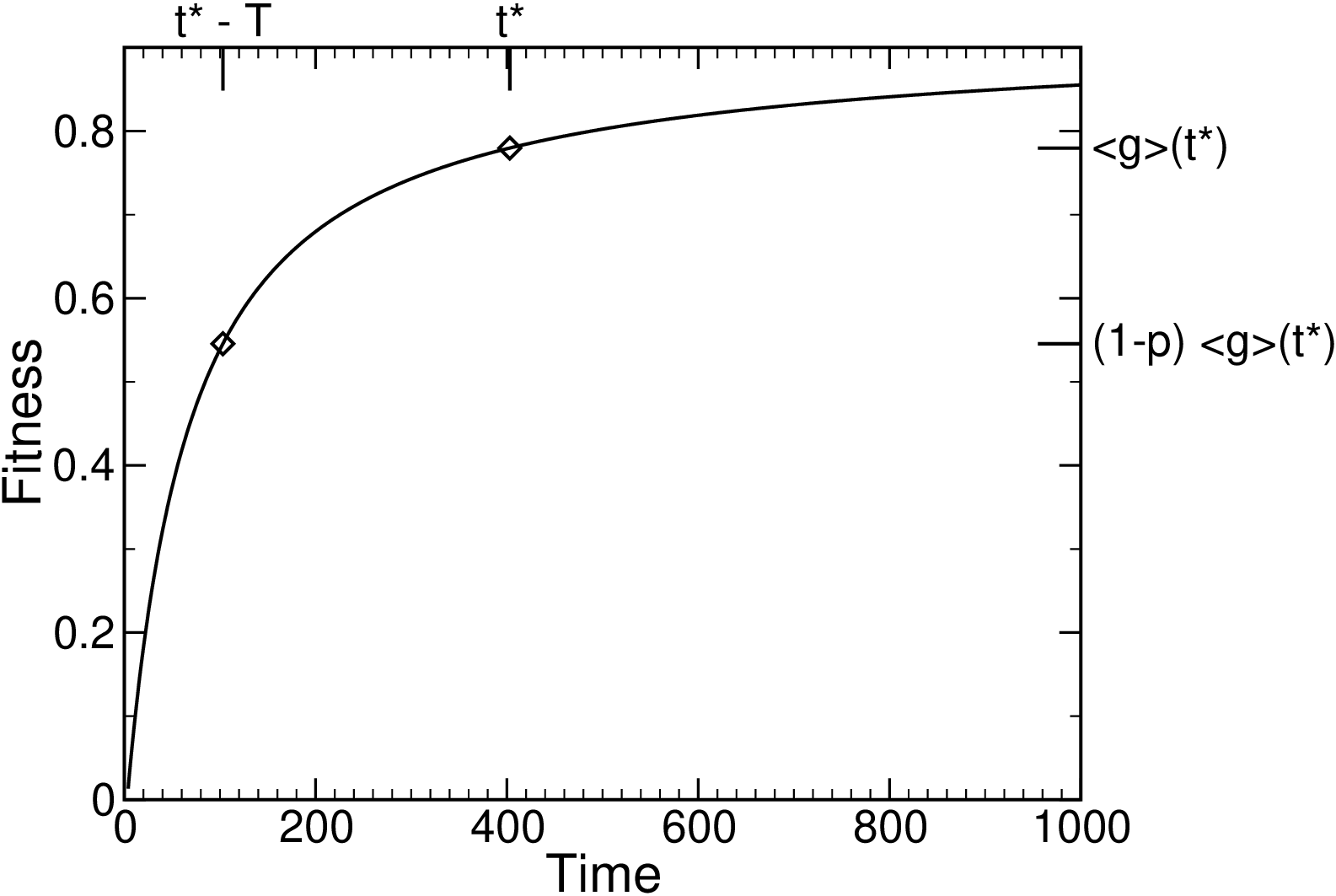}
\end{center}
\caption{a) Shown is the fitness of a single evolving system
with a given modularity  as a function
of time.
Positive fitness means growth of the system.
 The environment is repeatedly changed each $T=300$
time steps.  Shown in b) is the
average of these responses
 during a time $0$ to $T$ after each environmental change,
averaged over many environmental changes.
Shown in c) is the average response function to $p=1$ environmental
changes, $\langle g \rangle (t)$.
The response function in 
b) follows from a master response function curve in c),
being the
$t^* -T$ to $t^* $ subset where 
$\langle g \rangle (t^*-T) = (1-p) \langle g \rangle (t^*)$.
Here $p = 0.3$ and $T = 300$.
The present theory applies once the curve in c) has been determined.
}
\label{fig0}
\end{figure}

A changing environment will put pressure on
the system to have an efficient response function.
As the environment changes, the favorable niches
for the system change, and the system must adapt to the changing landscape.  The more rapidly the environment changes or
the more dramatically the environment changes, the more pressure there is 
on the system to be adaptable.  As noted above, it has been widely 
observed that systems under pressure tend to become more modular.  
The mean fitness of the systems a time $T$ after an
environmental change will depend
on the magnitude of the change, $p$, 
as well as the modularity.  We denote this value by
$f_{p, T}(M)$.
We can derive this function $f_{p, T}(M)$ for any $p$ and $T$ from the
average fitness as a function of time,  
starting from random initial conditions, 
which we denote as $\langle g \rangle(t)$, with
$\langle g \rangle(0) = 0$.
See Fig.\ \ref{fig0} for a depiction of the hierarchy of 
evolutionary timescales.
The observable $\langle g \rangle (t)$, Fig.\ \ref{fig0}c,
is an input to the theory
presented here and comes from
a detailed calculation, numerical simulation, or 
experimental observation of the competitive evolutionary dynamics.
The change of environment 
decreases the fitness by $1-p$ on average \cite{Earl}, 
and the time of evolution in each environment is $T$. 
These two conditions imply 
$f_{p, T}(M) = \langle g \rangle (t^*)$ where
$t^*$ is defined by
\begin{equation}
\langle g \rangle (t^*-T) = (1-p) \langle g \rangle (t^*)
\label{larget}
\end{equation}
The function $f_{p, T}(M)$ tells us the average, evolved
fitness of the system at the end of each environmental
change.   This function can be considered to be the fitness when
the environmental change is integrated out. 
This $f_{p, T}(M)$  is the fitness function that goes into Eq.\ (\ref{10}).

Evolution of modularity depends on how the response function
$f_{p, T}(M)$  of the system varies with the parameters
of environmental change, $p$ and $T$.
Since systems under stress tend to become more modular, 
an interpretation is that the average
fitness for a modular system is greater than that for a non-modular 
system, 
at least for small $T$ or large $p$ where stress is large.
This behavior has been observed in a model 
of systems evolving in a changing 
environment, when horizontal gene transfer is
included \cite{jun}.  
We have recently proved this canonical behavior for a Moran model of population
evolution in a glassy, modular fitness landscape \cite{Deem2014}.
Glassy evolutionary dynamics has been noted a number of times 
\cite{Sear,Goldenfeld2006}.
Conversely, at long time, the less modular system  
 should have a higher fitness,
because modularity is a constraint on the optima that can be achieved.

\begin{figure}[t!]
\begin{center}
\includegraphics[width=3in,clip=]{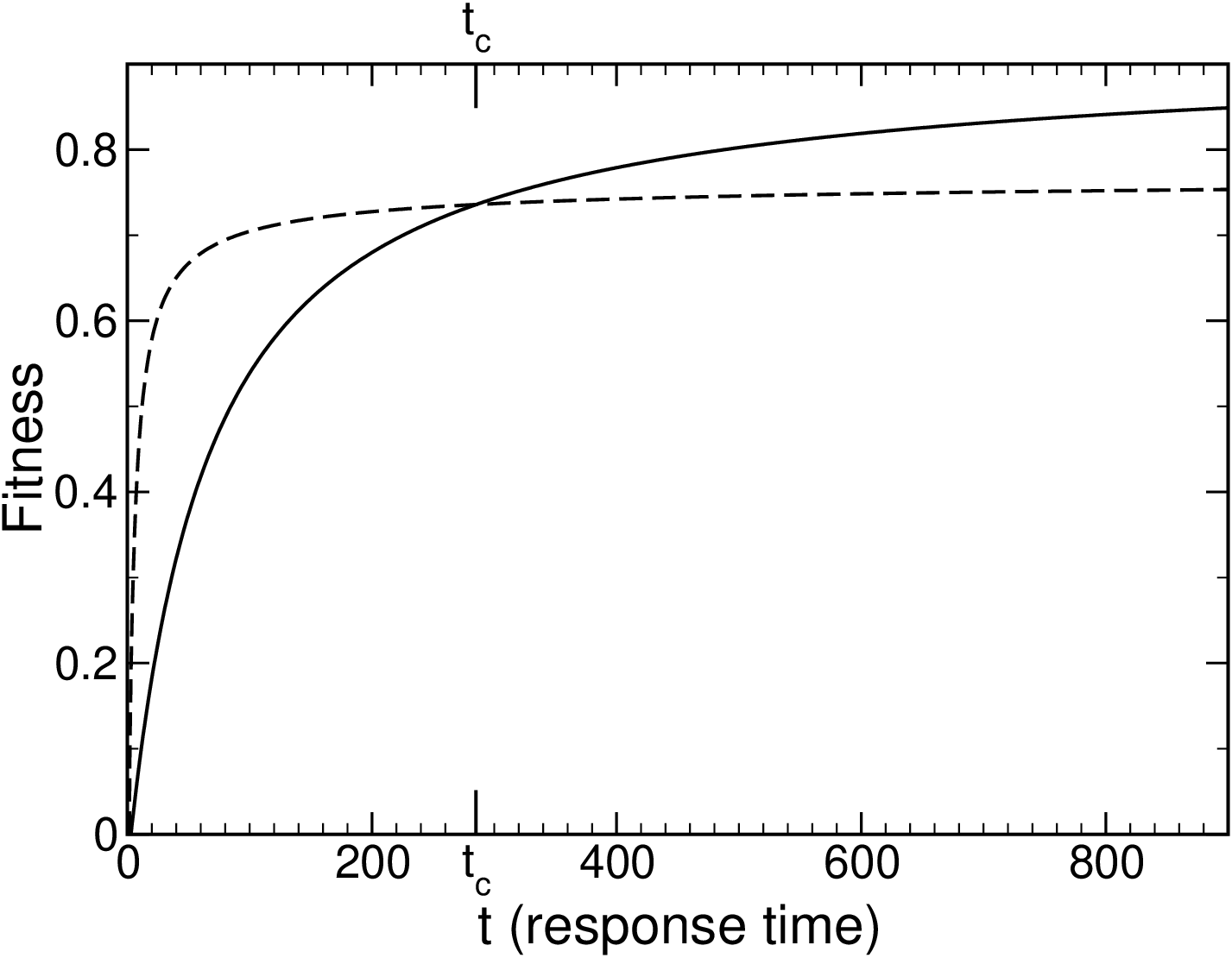}
\includegraphics[width=3in,clip=]{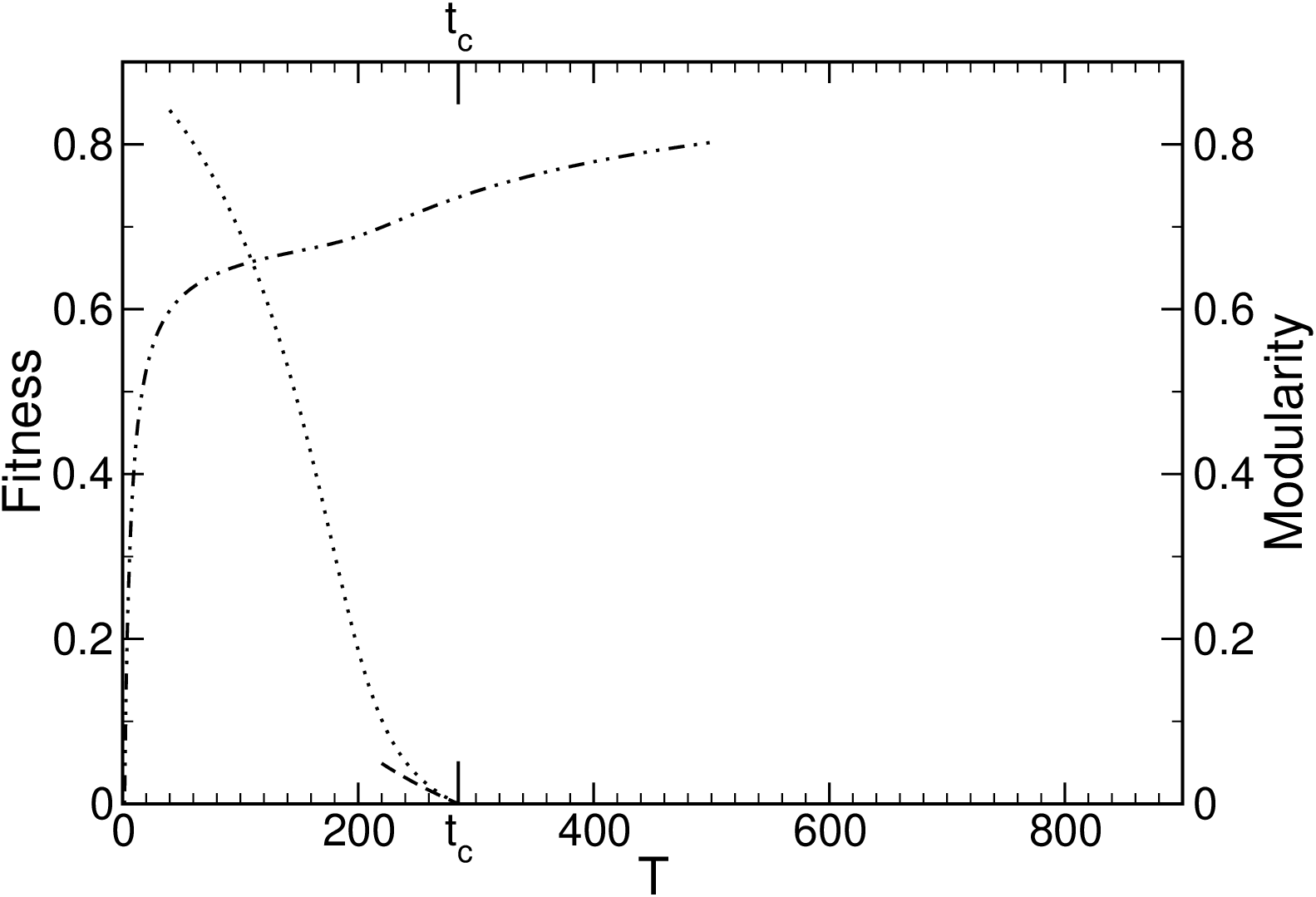}
\end{center}
\caption{Shown is the fitness of an evolving system. 
a) The fitness of the non-modular ($\langle g_0 \rangle$, solid)
and block-diagonal ($\langle g_1 \rangle$, dashed) system are shown,
starting from a random initial configuration.
These $\langle g_0 \rangle $ and $\langle g_1 \rangle$ are inputs to the theory.
The modular system is taken to be more fit at short time
and less fit at long time.
b) The evolved, steady-state
fitness of a system predicted
by the theory in a changing environment
(dot dashed), shown for 
varying $T$ and $p=1$.  The fitness follows the
high-modularity curve at rapid environmental changes, small $T$,
and the
low-modularity curve at slow environmental changes, large $T$.
Since $p=1$, the function $f_{p=1,T}(M) = \langle g(M) \rangle (t=T)$.  
The function $\langle g(M) \rangle $ is here
taken for simplicity to be $(1-M) \langle g_0 \rangle (t) + M \langle g_1 \rangle (t)$.
Note the modularity tends to 1 and the fitness to $\langle g_1 \rangle$ for rapid
environmental change (small $T$), and
the modularity tends to 0 and the fitness to $\langle g_0 \rangle$ for slow
environmental change (large $T$).
The modularity calculated from   theory, Eq.\ (\ref{3}), is shown
(dotted).  Also shown is the theoretical result for small $M$,
Eq.\ (\ref{4}), to first order in $l/L$ (short dashed).
In this example
$L=120$, $l=10$, $\mu = 0.01$, and $C=5.77$.
For these particular $\langle g_0 \rangle$ and $\langle g_1 \rangle$, the modularity emerges only for environmental
changes that occur on a timescale $T < t_c \approx 285$.
}
\label{fig2}
\end{figure}

In Eq.\ (\ref{10}), we here take this function  $f(m)$ as input.  
We assume only that    
the population averages for large $M$ and small $M$ look like the dashed and 
solid curves in Fig.\ \ref{fig2}a.
Putting these points together, the  quasispecies theory
presented here quantitatively describes
the emergence of
modularity at small $p$ or large $T$, as shown in
Figs.\ \ref{fig1} and
 \ref{fig2}b.

\section{A Fluctuation Dissipation Theorem}
There is a fluctuation dissipation relation
for the rate of change of modularity.
Multiplying Eq.\ (\ref{10}) by $m$ and summing, we find
that the rate of change of modularity satisfies
\begin{equation}
\frac{d M}{d t} = L \langle m f(m) \rangle - L M \langle f \rangle - \mu M
\label{11}
\end{equation}
This equation is a type of continuous-time Price equation \cite{Price}.
This equation implies a type of useful fluctuation-dissipation theorem.
Expanding $f(m)$,
we can alternatively write this 
fluctuation dissipation relation
describing the evolution of modularity as
\begin{equation}
\frac{d M}{d t}  \approx L \left. \frac{ d f} {d m} \right\vert_M
\langle \sigma^2_M \rangle - \mu M
\label{11a}
\end{equation}
Here $M = \langle m \rangle$ is the average modularity of the system,
and $\sigma^2_M = \langle m^2 \rangle - M^2$ is the variance of the
modularity, where
$m$ is the modularity for any particular system in the population.

\section{Environmental Change Selects for Modularity}

We now derive a
relationship between the rate of growth of modularity and
the environmental pressure.  
We investigate the dynamics for small modularity, and
 we consider a Taylor series expansion of the fitness function:
$f(m) = f(0) + m \Delta f +  o(m)$.
The function $\Delta f$ is time independent, depending on $p$, 
$T$, and other parameters of the evolution within each system that
have been projected out. 
We investigate the growth of modularity from an
initially non-modular state.
We consider how the response function depends on $p$.
If $p=0$, the environment is not changing, 
$t^* \to \infty$ in the expression of Eq.\ (\ref{larget}),
and the system 
will stay in the $M=0$ state.  This implies $\Delta f = 0$ when
$p=0$, as otherwise
a non-zero modularity would emerge, see Eq.\ (\ref{4}) below.
For small $p$, the environment is changing only slightly,
$t^*$ is large,
 and the
system will evolve a small value of $M$.
Expanding in a Taylor series for small $p$ and $T \ll t^*$,
Eq.\ (\ref{larget}) becomes
\begin{equation}
\frac{p}{T} = 
\frac{\langle g_m' \rangle (t_m^*) }{ \langle g_m \rangle (t_m^*) }
\approx \frac{ \langle g_m' \rangle (t_m^*)}{ \langle g_m \rangle (\infty) }
\approx \frac{\langle g_m' \rangle(t_m^*) }{ \langle g_0 \rangle(\infty)}
\label{taylor2}
\end{equation}
where the last two relationships arise
because 
$\langle g_m' \rangle (t_m^*)$ is small and because
$t_m^*$ is large  and  $m$ is small.
Thus,  $\Delta f = \lim_{m \to 0} [f(m) - f(0)]/m = \Delta f(p/T)$.
Expanding $\Delta f$ to first order in $p/T$ and taking $m$ small, we find
$\Delta f = \alpha p/T$.
When $m$ is small, equation (\ref{11}) becomes
\begin{equation}
M' = L \sigma^2_M \Delta f - \mu M
\label{3c}
\end{equation}
Using the result above for $\Delta f$, 
we find $M' \approx  \alpha L \sigma^2_M p/T$,
leaving out the small term
proportional to $M$ in Eq.\  (\ref{3c}).
We, thus, find
\begin{equation}
p_{\rm E} \approx \frac{1}{R}\frac{d M}{ d t}
\label{13}
\end{equation}
where $p_{\rm E} = p/T$ is the environmental pressure, and
$R =  \alpha L \sigma_M^2  $.
In this equation,
$R  \propto \langle \sigma_M^2 \rangle$, which as experimentalists
have anticipated is related to
replicate variability in experiments \cite{Cooper}.

This Eq.\ (\ref{13})  follows from the 
fluctuation dissipation relation
in Eq.\ (\ref{11}) and
the response function of the modular system being
greater than that of the non-modular system at short time.
Equation (\ref{13}) 
may be interpreted as a Taylor series
expansion of $dM/dt$ in allowed combinations of
$p$ and $1/T$. Alternatively, Eq.\ (\ref{13})
may be interpreted as the linear response of the modularity
to the environmental pressure.
The coefficient $R$ is a measure of ruggedness of the 
evolutionary landscape within each system.
This ruggedness slows down the evolutionary
dynamics, and the selection for an effective response function provided
by a changing environment implicitly selects for modularity when 
horizontal gene transfer is active \cite{jun}.
Here, we are able to show that
$R$ is proportional to the variance
of the modularity, which is expected to be related to the ruggedness
of the landscape.
It is the ruggeddness of the landscape that leads to non-trivial replicate
variability.

For what forms of
$\langle g_m \rangle (t)$ will the
$\Delta f(p/T)$ function be analytic in $p/T$? 
We first consider an exponential convergence of the
fitness function: 
$\langle g_m \rangle (t) = g(\infty) - a_m \exp (-\beta_m t)$,
where we have left out the $m$ depencence of  $g(\infty)$ because
we expect it to be higher order than linear in $m$.
Eq.\ (\ref{taylor2}) becomes $p g(\infty)/T = a_m \beta_m \exp (-\beta_m t^*_m)$,
and we find
$ f_{p,T} (m) = \langle g_m \rangle (t^*_m)  = g(\infty) - g(\infty) p / (T \beta_m)$.
We thus find 
 $m \Delta f = g(\infty) (p / T) (1/\beta_0 - 1/\beta_m)$, which
is positive because we expect the modular system to converge
faster, $\beta_m > \beta_0$.
Thus, 
we find Eq.\ (\ref{13}), with $\alpha = -g(\infty) d \beta_m^{-1} / d m
 \vert_{m=0}$.
Conversely, for a power law decay
$\langle g_m \rangle (t) = g(\infty) - a_m t^{- \beta}$,
we find the fitness to be non-linear in $p/T$:
$ f_{p,T} (m)  = g(\infty) - a_m [ (p g(\infty) / (T a_m \beta)]
^{\beta/(\beta+1)}$.
In this case, Eq.\ (\ref{13}) is modified 
to be $p_{\rm E} ^{\beta/(\beta+1)}$ on the left hand side,
with $\alpha = - [g(\infty)/\beta]^{\beta/(\beta+1)} 
d a_m^{1/(\beta + 1)} / d m \vert_{m=0}$.
Finally, for a logarithmic decay \cite{Deem2014}
$\langle g_m \rangle (t) = g(\infty) - a_m  \ln^{-2/\nu} (t / t^0_m)$,
we find the fitness to be non-analytic in $p/T$,
since  $(p/T) t^*_m   g(\infty) = (2/\nu) a_m \ln^{-2/\nu-1} (t^*_m / t^0_m)$.
This equation can be solved in terms of powers of the
product logarithm, or Lambert $W_0$ function.
Performing an asymptotic analysis for small $p/T$, we
find  $ f_{p,T} (m)  \sim g(\infty) - a_m \ln^{-2/\nu} (T/p)$.
In this case,  Eq.\ (\ref{13}) is modified
to be $1/(\ln^2 p_{\rm E})^{1/\nu}$ on the left hand side, with
$\alpha = - d a_m / d m \vert_{m=0}$.

Equation (\ref{13}) is a description of how the evolvability
of the system depends on the environmental change.
That is, $d M / d t$ is a measure of the evolvability of the system,
with larger values indicating a greater rate of change
of the measurable order parameter $M$.  This measure
of evolvability is greater for greater environmental pressures,
$p_{\rm E}$.  The drive for spontaneous emergence of modularity,
large $dM/dt$, 
is also greater for landscapes that are more rugged, i.e. larger $R$,
which can be estimated from variability of replicate experiments.

Equation (\ref{13}) says that an increase of
environmental pressure should
lead to the evolution of systems with increased modularity. 
A study of 117 species of bacteria showed 
that the modularity of the bacteria's metabolic networks 
increased monotonically with variability of the
environment in which the bacteria lived \cite{parter2007}.
Metabolic networks of
pathogens alternating between hosts were found to be
more modular than those of single-host pathogens \cite{kreimer2008}.

\section{Steady-State Values of Modularity in One Environment}

\subsection{Field Theory for the Dynamics of Modularity}

Here we rewrite the dynamical equations of quasispecies
theory in the language of field theory.
We solve the field theory in the limit of large system sizes
to determine the steady-state modularity that emerges at long time.
The theory is distinct from traditional quasispecies theory because
the replication rate depends on the modularity rather than
the Hamming distance from a wild-type strain.  Nonetheless, 
we will show that the theory
can still be solved exactly in the limit of a large system size.

For large values of $L$, for which the changes in $M$ are nearly continuous, 
we here determine
the average fitness implied by Eq.\ (\ref{10}) at long time
by  techniques borrowed from quantum field theory
 \cite{Peliti2002,Park06}.
We write the dynamical equations in 
 Eq.\ (\ref{10}) in terms of raising and lowering operators.
We then use coherent states to write this second quantization
in terms of a Bosonic field theory, with fields $z^*_{ij}(t),
z_{ij}(t)$ representing density at $\Delta_{ij}(t)$
at time $t$.
The action of this field theory is
\begin{eqnarray}
S[\{ z \}, \{ z^* \}] &=&
\int_0^{t_f} \sum_{ij}
z_{ij}^* (t)  \partial_t z_{ij} (t) dt
+ \sum_{ij} 
\left[ 
z_{ij}^* (0)  z_{ij} (0) 
-
z_{ij}(t_f) 
\right] + L C
\nonumber \\
&&
- \frac{C_1}{L} \sum_{ij}^{\rm in} z_{ij}^*(0)
- \frac{C_0}{L} \sum_{ij}^{\rm out } z_{ij}^*(0)
\nonumber \\ &&
- L \int_0^{t_f} f \left[
\frac{1}{L C}
\left( \sum_{ij}^{\rm in} z_{ij}^*(t) z_{ij}(t)
- \frac{1}{L/l-1}
 \sum_{ij}^{\rm out} z_{ij}^*(t) z_{ij}(t)
\right)
\right] dt
\nonumber \\ &&
- \frac{\mu}{L^2} \int_0^{t_f} \sum_{ij} \sum_{mn} [z_{mn}^*(t) - 
z_{ij}^*(t)] z_{ij}(t) dt
\label{12b}
\end{eqnarray}
Note that the fitness depends on the modularity of
the connection matrices of each state at each point in 
time in Eq.\ (\ref{12b}), just as it
did in Eq.\ (\ref{10}).  Also note that Eqs. (\ref{10}) and (\ref{12b}) are
exact for arbitrary, non-linear fitness functions $f(m)$.
Here ``in'' means in the $l \times l$ block diagonals and
``out'' means outside these block diagonals.
The quadratic terms can be integrated out (see Appendix B) \cite{Park06},
and we are left with an action
expressed in terms of a modularity field, $\xi$, and its conjugate, $\bar \xi$:
\begin{eqnarray}
S &=&
L \int_0^{t_f} [C \bar \xi(t) \xi(t) -  f(\xi(t)) ] dt
-L C \ln Q
\label{12c}
\end{eqnarray}
where the
determinant is 
$Q = [l C_1(t_f) + (L-l) C_0(t_f)] / (L C)$, where the vector
${\bf C}(t) = (C_1(t), C_0(t))$ satisfies 
\begin{equation}
d {\bf C} / dt = A(t) {\bf C} (t)
\label{12e}
\end{equation}
where \begin{equation}
A(t) = 
\left(
\begin{array}{cc}
- \mu (L-l)/L + \bar \xi(t) & \mu (L - l)/L\\
\mu  l / L & -\mu l / L - \bar \xi(t) l / (L-l) 
\end{array}
\right)
\label{12d}
\end{equation}
 and ${\bf C}(0) = (C_1, C_0)$.

\subsection{The Steady-State, Average Value of Modularity}
The average modularity follows 
a dynamical trajectory away from an initial state to
a final steady state value.
For large $L$, 
this action becomes large, and a saddle point
calculation can be used
(see Appendix C).
The remarkable result from this derivation is that the modularity which
emerges at long time obeys a principle of least action:
\begin{eqnarray}
f_{\rm pop} = \max_\xi \left\{ f(\xi) - \mu C [(L-l) l / L^2]
[2 + (L/l-2) \xi - 2 \sqrt{
(1-\xi) (1 + (L/l-1) \xi)
}
]
\right\}
\label{3}
\end{eqnarray}
The variance of the modularity is small, ${\cal O}(1/L)$, and 
the modularity is determined by the solution of the
implicit equation 
\begin{equation}
f(M) = f_{\rm pop}
\label{3a}
\end{equation}
Here $f_{\rm pop}$ is the mean population fitness, i.e.\ Eq.\ (\ref{3b}) with
$t \to \infty$.
Thus, a principle of least action gives the evolved modularity
at steady state.
Coexistence of populations with different modularity, i.e.\ bimodality
in the distribution of modularity, is possible if the $f(m)$ function
is discontinuous \cite{Park}.

\subsection{Phase Diagrams for The Emergence of Modularity}
While Eq.\ (\ref{3}) is a general result, we can proceed further in 
the limit that evolved modularities are small.
Expanding for small $M$, we find
\begin{eqnarray}
\xi_{\rm max} &\sim& \frac{2 l 
\left[ d f/dM \vert_{M=0} \right] }{ \mu C (L-l) - 2 l 
\left[ d^2 f/d M^2 \vert_{M=0} \right] }
\nonumber \\
f_{\rm pop} &\sim& \frac{l \left[d f/dM \vert_{M=0}\right]^2}{ \mu C (L - l) - 2 l 
\left[ d^2 f/dM^2 \vert_{M=0} \right] } + f(0)
\nonumber \\
M &\sim& \frac{l \left[ d f /dM \vert_{M=0} \right]}{\mu C (L-l) - 2 l 
\left[d^2 f/dM^2 \vert_{M=0} \right]}
\label{4}
\end{eqnarray}
Thus, as long as a modular system has a higher fitness, $df/dM > 0$,
modularity will spontaneously emerge, $M > 0$,
for large enough system sizes, $L$.  
Note also when $M$ is small, that the steady state modularity
calculated exactly from Eq.\ (\ref{3})  is in agreement with the
small $M$ result in Eq.\ (\ref{4}), as shown in 
Fig.\ \ref{fig2}b.
Note that for large $L/l$, Eq.\ (\ref{3c}) combined with Eq.\ (\ref{4})
 implies that at steady state
$ \sigma^2_{M_\infty} = l/[L(L-l) C]$.

\begin{figure}[t!]
\begin{center}
\includegraphics[width=3in,clip=]{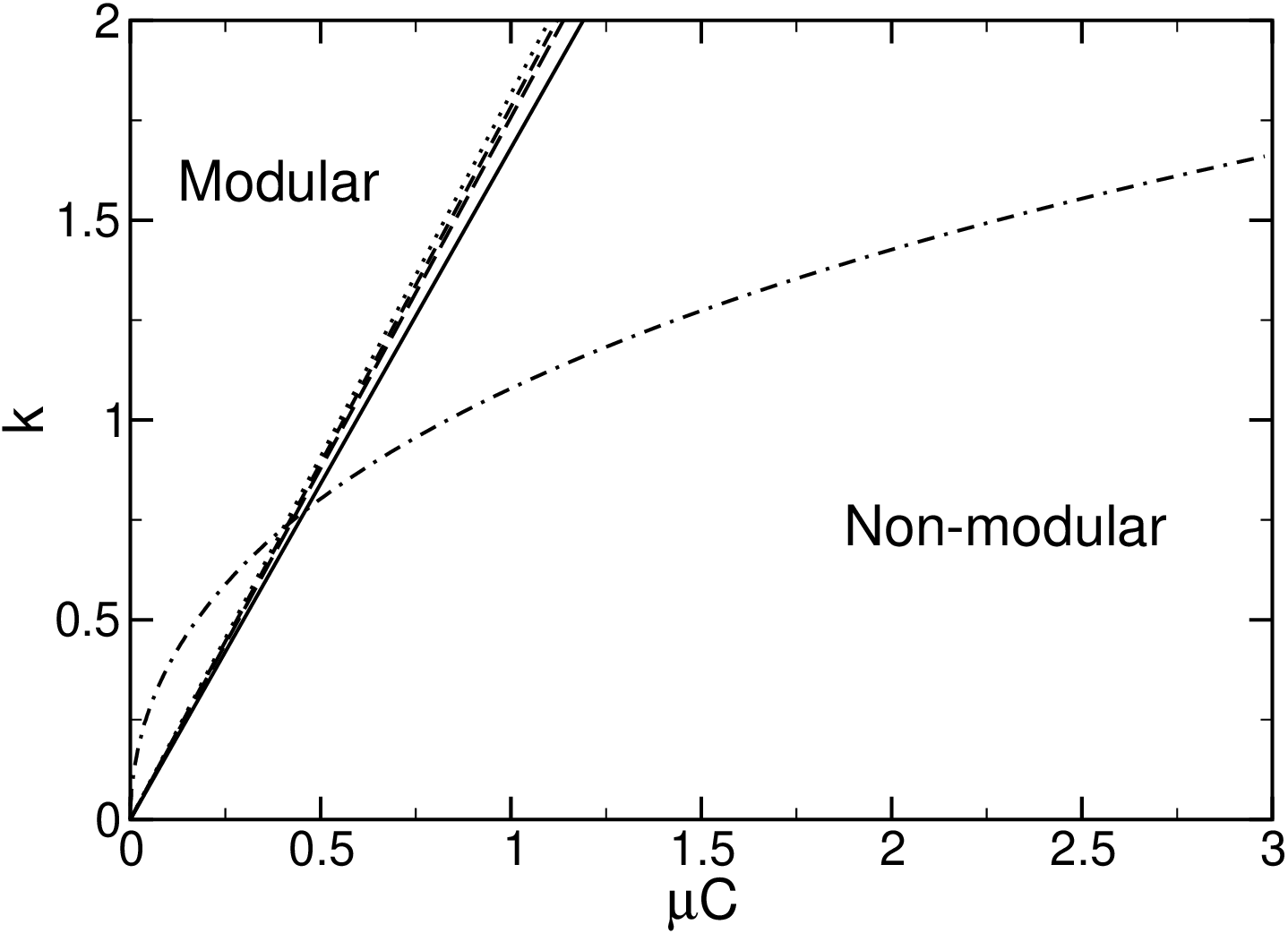}
\end{center}
\caption{
The phase diagram for emergence of modularity.
Below a critical mutation rate, modularity spontaneously
emerges.  Results are shown for $f(M) = k M^2/2$ (solid),
$f(M) = k M^3/2$ (long-dashed),
$f(M) = k M^4/2$ (short-dashed), 
$f(M) = k M^{10}/2$ (dotted), and
$f(M) = e^{k M} - k M - 1$ (dot-dashed).
Results here are shown for $l = 10, L = 120$.
}
\label{fig1}
\end{figure}

For fitness functions
for which $df/dM \vert_{M=0} = 0$, more analysis is required.
For example, if $f(M) = k M^2/2$, there is a phase transition at
$\mu^* $:  For $\mu < \mu^*$ modularity emerges,
whereas for $\mu > \mu^*$ the population remains in the non-modular phase.
This phase transition is analogous to the error catastrophe found
in traditional quasispecies theory.
Phase diagrams for a number of fitness functions are
shown in Fig.\ \ref{fig1}.

\section{Using Quasispecies Theory to Extrapolate Simulation Data
on Spontaneous Emergence of Modularity}

\begin{figure}[t!]
\begin{center}
\includegraphics[width=3in,clip=]{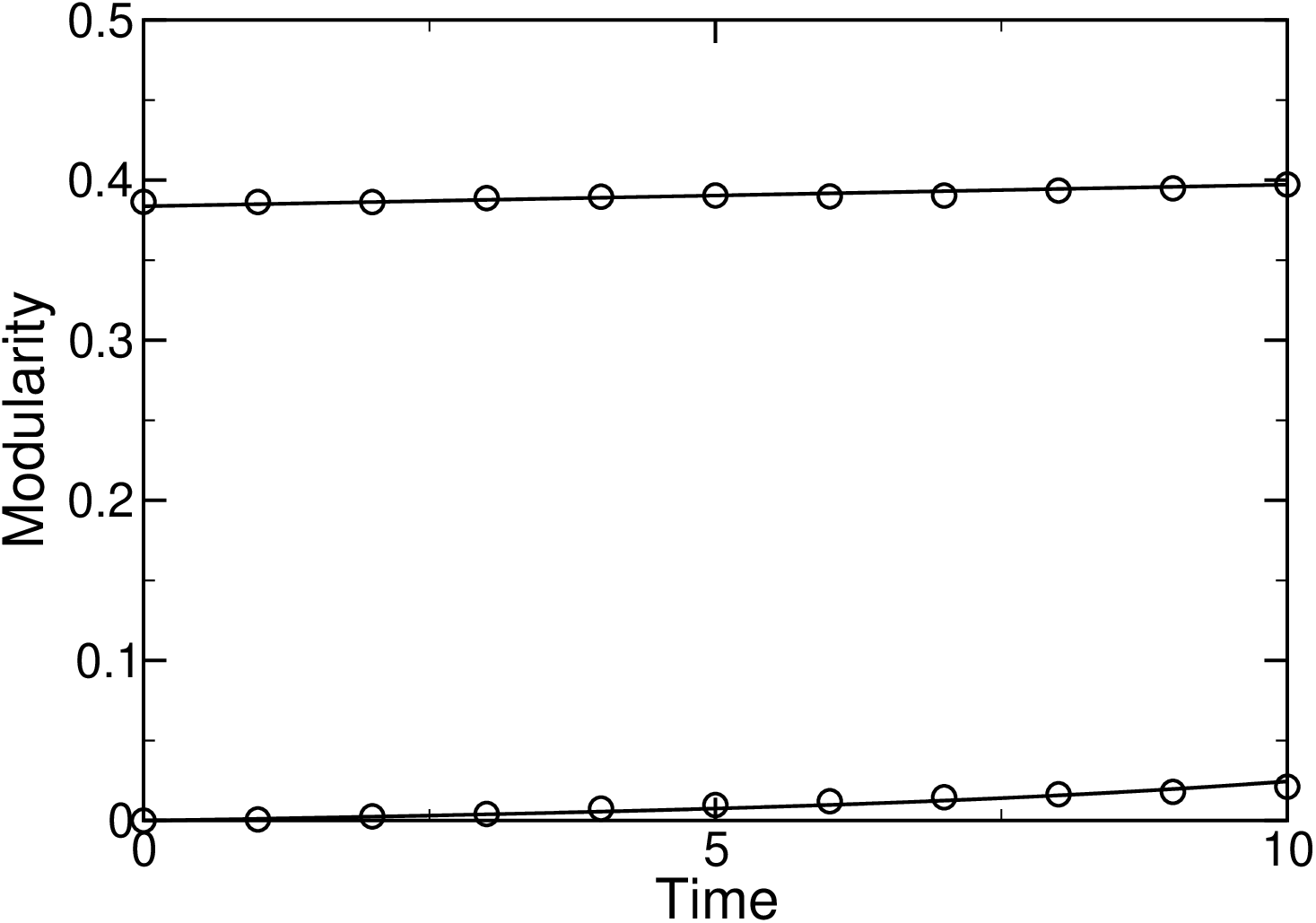}
\end{center}
\caption{Shown is modularity versus time for a population that
exhibits spontaneous emergence of modularity.
The curves are from theory, Eq.\ (\ref{10}), and the
data (circles) are from \cite{jun}.
Two different initial conditions are shown, $M(0) = 0$ and
$M(0) = 0.38$.
In this example the derived underlying fitness function
is $f(M) = 1.4 M - 1.31 M^2$, the mutation rate
is $\mu = 2/346$, and the average number of connections is
$C = 346 \times 2 / L = 5.77$.
}
\label{fig3}
\end{figure}

We use Eq.\ (\ref{10}) to analyze $M(t)$ 
data on spontaneous emergence of modularity in a simulation
of an evolving protein network \cite{jun}
to deduce $df/dm$ and to derive $f(M)$ by
integration.
For this system, we know the mutation rate,
as two of the connections change per time step in the upper half
of the connection matrix,
and so we can use Eq.\ (\ref{10}) at short time 
to determine $df/dm$.  Alternatively we can
determine $df/dm$ if we know the
variance of the modularity and $M(t)$,
c.f.\ Eq.\ (\ref{11a}).
We assume $f(M)$ is quadratic, and integrate the $df/dm$ 
to determine the $f(M)$.
There are $N_D = 346$ total connections in the upper half of the
connection matrix and $N_0 = 22$ connections in the upper half  of the
connection matrix when $M=0$ for the parameters of \cite{jun}.
Thus, we take $C = 346 \times 2 / L = 5.77$ and
$\mu = 2/346$.
When $M=0$, the population was prepared by four discrete time
iterations of the mutation step, from a single initial
configuration \cite{jun}.  We find $f(M) \sim 1.4 M$ reproduces the data
at small $M$.
For the initial condition of $M = 0.38$, the configurations
were taken from an ensemble \cite{jun}, 
which we take to satisfy Eq.\ (\ref{10}).
We find $f(M) = 1.4 M - 1.31 M^2$ approximately reproduces the data,
 as shown in Fig.\ \ref{fig3}.
Equation (\ref{3}) predicts
a steady-state value of $M= 0.45$, toward which the computationally
costly simulations appear to be heading.

\section{Conclusion}
The examples of environmental stress leading to modularity,
ranging from metabolic networks of bacteria in different physical
environment to simulations of emergence of protein secondary structure,
can be quantified by quasispecies theory.
The approximate relation $R p_{\rm E} = dM /dt$ relates rate of
growth of modularity to the ruggedness of the fitness landscape, $R$,
and environmental pressure, $p_{\rm E}$, for small values of modularity.
The present theory should allow the analysis of
complex, evolving populations to go beyond a demonstration of the
existence of modularity to 
a quantitative analysis of the dynamics of modularity.
That is, the theory presented here should allow the determination of the
$f(M)$ function for these evolving populations, by using the predictions
to determine the $f(M)$  that best matches observation.
Knowing the $f(M)$ and $\mu$ that fundamentally characterize a population 
would then allow for out-of-sample predictions of dynamical modularity.

\section*{Acknowledgments}
This research was supported by US National Institutes of
Health, 1 R01 GM 100468--01.
JMP was also supported by the Catholic University of Korea research
fund 2014 and by the National Research Foundation of Korea Grant
(NRF-2011-013-C00029 and NRF-2013R1A1A2006983).

\section{Appendix A}
We here derive Eq.\ (\ref{10}).  The rate to 
 increase modularity for a matrix with modularity $m$ is
$r_{\rm up} = \mu n_{\rm out} (L/l)  l^2/L^2$.
Recall we are in the dilute limit: $C$ is finite, and $L$ is large.
Thus, collisions between entries in the connection matrix can be ignored.
The rate  to
 decrease modularity for a matrix with modularity $m$ is
$r_{\rm down} = \mu n_{\rm in}  L (L-l)/L^2$.
Here the number of connections inside the $l \times l$ blocks is
given by $n_{\rm in}$ and
the number of connections outside the $l \times l$ blocks is
given by $n_{\rm out}$.
We have the constraint $n_{\rm in} + n_{\rm out} = C L $.
We also have by the definition of modularity
$m = [ n_{\rm in}/l - n_{\rm out}/ (L-l)] l / (C L)$,
which shows modularity changes by discrete  increments
of $\pm 1/[ C (L-l)]$.
Thus, we find
$r_{\rm up}(m) = \mu C l (L-l) (1-m)/L$
and
$r_{\rm down}(m) = \mu C  (L-l) (L m - l m + l)/L$.
For non-zero modularity, to avoid collisions in the $\Delta$ matrix, we further 
require $\langle n_{\rm in} \rangle \ll  l L$, 
i.e.\ $C (l - l M + L M) \ll l L$.
Alternatively, if this constraint is not satisfied, we can view
Eq.\ (\ref{10}) as a generalization to the case of integer occupation
numbers of the $\Delta$ matrix with certain biased hopping probabilities,
$r_{\rm up}(m)$ and $r_{\rm down}(m)$, given above.
The rate of change of $P_m(t)$ due to replication
is $L [f (m)- \langle f \rangle] P_m(t)$,
where the second term ensures conservation of probability,
$\sum_m P_m(t) = 1 \  \forall \  t$.
This is the first term on the right hand side in Eq.\ (\ref{10}).
The rate of increasing $P_m(t)$ due to an increase
of modularity from $m - 1/[C (L-l)]$ to $m$ is
$r_{\rm up}[m - 1/( C (L-l))] P_{m - 1/[C (L-l)]}(t)$,
which is the first $\mu$-dependent term in Eq.\ (\ref{10}).
The rate of increasing $P_m(t)$ due to a decrease
of modularity from $m + 1/[ C (L-l)]$ to $m$ is
$r_{\rm down}[m + 1/(C(L-l))] P_{m + 1/[C (L-l)]}(t)$,
which is the second $\mu$-dependent term in Eq.\ (\ref{10}).
The rate of decreasing $P_m(t)$ due to modularity
changing from $m$ to $m \pm 1/[C (L-l)]$ is
$[r_{\rm up } (m) + r_{\rm down } (m) ]
 P_{m }(t)$,
which is the third $\mu$-dependent term in Eq.\ (\ref{10}).
Thus, we have derived  Eq.\ (\ref{10}).

\section{Appendix B}
We here calculate the determinant
that comes from integrating out the $z^*_{ij}$ and $z_{ij}$
fields in Eq.\ (\ref{12b}).  
The probability of connections inside
and outside the blocks have been taken initially
to be Poisson in Eq.\ (\ref{12b}), with average probability
of a connection per site to be $C_1/L$ inside the blocks
and $C_0/L$ outside the blocks.
The overall average number of connections per row is  $C = C_0 + (C_1 - C_0)l/L$.
We here project the number of connections 
onto the constraint that there are
$L C$ total connections.   
As in \cite{Park06}, this constraint is
enforced with a projection operator that leads to twisted
boundary conditions.  
A modularity field $\xi$ and conjugate field $\bar \xi$ are defined,
with $\xi(t)$ as the argument of the fitness function  in
Eq.\ (\ref{12b}).
We use a trotter factorization and
define $\epsilon = t_f/M$ and will take the limit $M \to \infty$.
We define $\delta = 1$ if 
$\lfloor i/l \rfloor
= 
\lfloor j/l \rfloor$ and zero otherwise.
The partition function becomes
\begin{eqnarray}
\cal Z &=& \int 
[{\cal D} \bar \xi {\cal D} \xi ]
e^{
-\epsilon L C \sum_{k=1}^M \bar \xi (k) \xi(k) +
\epsilon L \sum_{k=1}^M f[\xi(k)]
}
\nonumber \\
&& \times
\int_0^{2 \pi} \frac{d \eta}{2 \pi} e^{-i \eta - L C}
[{\cal D} z^* {\cal D} z]
e^{
-\sum_{k=0}^M \sum_{ij} z^*_{ij} (k) z_{ij} (k) +
\sum_{ij} z_{ij} (M) 
}
\nonumber \\ && \times
e^{
\sum_{k=1}^M \sum_{ij} \left[
z^*_{ij} (k) 
+
(\epsilon \mu /L^2) \sum_{mn}(z^*_{mn}(k) - 
z^*_{ij}(k) ) 
+ \epsilon \bar \xi(k) (L \delta - l)/(L-l) z^*_{ij} (k) 
\right]
z_{ij}(k-1)
}
\nonumber \\ && \times
e^{
(C_1(0)/L) e^{i \eta / (L C)}  \sum_{ij}^{\rm in} z^*_{ij}(0)
+ (C_0(0)/L) e^{i \eta / (L C)}  \sum_{ij}^{\rm out} z^*_{ij}(0)
}
\label{A10}
\end{eqnarray}
Integrating out $z^*_{ij}(0)$ and $z_{ij}(0)$, the action remains the
same except the start on sums over $k$ are incremented by one, and the
terms $C_1(0) z^*(0)$ and $C_0(0) z^*(0)$ 
become
$C_1(1) z^*(1)$ and $C_0(1) z^*(1)$ 
with
\begin{eqnarray}
C_1(1) &=& C_1(0) \left[ 1 - \epsilon \mu \left(1-\frac{l}{L}\right) 
+ \epsilon \bar \xi(1)\right] + C_0(0) \epsilon \mu \left(1-\frac{l}{L}\right)
\nonumber \\
C_0(1) &=& C_0(0) \left[ 1 - \epsilon \mu \frac{l}{L} 
- \epsilon \bar \xi(1) \frac{l}{L-l}\right] 
+ C_1(0) \epsilon \mu \frac{l}{L}
\label{A11}
\end{eqnarray}
Iterating the process of integrating out the $z^*(k)$ and $z(k)$, we find
that the
vector ${\bf C}(t) = (C_1(t), C_0(t))$ renormalizes according
to Eq.\ (\ref{12e}).
Finally, integrating out 
$z^*(M)$ and $z(M)$, we find
the final contribution to the partition function is
\begin{eqnarray}
\cal Z &=& \int 
[{\cal D} \bar \xi {\cal D} \xi ]
e^{
-\epsilon L C \sum_{k=1}^M \bar \xi (k) \xi(k) +
\epsilon L \sum_{k=1}^M f[\xi(k)]
}
\nonumber \\
&& \times
\int_0^{2 \pi} \frac{d \eta}{2 \pi} e^{-i \eta - L C}
e^{ [ l C_1(M) + (L-l) C_0(M)] e^{i \eta / (LC)}
}
\label{A12}
\end{eqnarray}
Performing the final integration over $\eta$, we find
the final expression for the partition function to be
\begin{eqnarray}
\cal Z &=& \int 
[{\cal D} \bar \xi {\cal D} \xi ]
e^{
-\epsilon L C \sum_{k=1}^M \bar \xi (k) \xi(k) +
\epsilon L \sum_{k=1}^M f[\xi(k)]
}
\left[
\frac{l C_1(M) + (L-l) C_0(M)}
{LC}
\right]^{L C}
\label{A13}
\end{eqnarray}
Thus, the action in Eq.\ (\ref{12c})
is derived.

\section{Appendix C}
Here we calculate the saddle-point solution to the action (\ref{12c})
at large time.
For large $L$, this saddle point solution is exact.
For large $t_f$, Eq.\ (\ref{12c}) becomes
\begin{equation}
S = L t_f [C \bar \xi \xi -  f(\xi) ]
-L C \ln Q
\label{A1}
\end{equation}
where
\begin{equation}
Q = Tr \left[ e^{t_f A}
\left(
\begin{array}{c}
C_1(0) / C\\
C_0(0)/C
\end{array}
\right)
\left(
\frac{l}{L} 
,
\frac{L - l}{L}
\right)
\right]
\label{A2}
\end{equation}
The larger eigenvalue of $A$ is given by
\begin{equation}
\lambda_+ = -\frac{1}{2}\left( \mu - \frac{L-2 l}{L-l} \bar \xi \right)
+ \frac{1}{2} \left[
\left(\mu - \frac{L - 2l }{L-l} \bar \xi \right)^2 + \frac{4 l \bar \xi^2}{L - l}
\right]^{1/2}
\label{A3}
\end{equation}
Thus, the action tends to
\begin{equation}
-S \sim L t_f [-C \bar \xi \xi +  f(\xi) ]
+L C t_f \lambda_+
\label{A4}
\end{equation}
Maximizing this over $\bar \xi$, we find
\begin{eqnarray}
-S/(L t_f) \sim f(\xi) - \mu C [(L-l) l / L^2]
[2 + (L/l-2) \xi - 2 \sqrt{
(1-\xi) (1 + (L/l-1) \xi)
]
}
\label{A5}
\end{eqnarray}
Maximizing over $\xi$ gives Eq.\ (\ref{3}).  Using that the partition
function $\cal Z$ grows at long time as $\exp(L f_{\rm pop} t_f)$ \cite{Park06},
we find Eq.\ (\ref{3a}).

\bibliography{pE}

\end{document}